# Quantum walk in (1+1)-dimensional spacetime for Majorana dynamics with high order approximation in NISQ

Wei-Ting Wang[1], Xiao-Gang He[1,2], Hsien-Chung Kao[3], and Ching-Ray Chang[4]

[1]*Department of Physics, National Taiwan University, Taiwan*
[2]*Tsung-Dao Lee Institute, Shanghai Jiao Tong University, China*
[3]*Physics Department, National Taiwan Normal University, Taiwan*
[4]*Department of Physics and Quantum Information Center, Chung Yuan Christian University, Taiwan*

In this study, we show that quantum walk can describe a Majorana fermion when the coin operator constrained by Lorentz covariance and the initial state satisfies the Majorana condition. The time evolution of a Majorana fermion is demonstrated with the numerical simulations and experimentally runs on a real quantum device provided by IBM Quantum System. To reduce errors due to approximation, we proposed a new efficient way to achieve second order accuracy in the near-term quantum computer without increase the complexity of quantum gate circuitry compared with the first order approximation. We show that rest Majorana fermion (expectation value of momentum is zero) can be well defined and its behavior depends more sensitively on the accuracy of the approximation than a Dirac particle due to the stringent constraints of Majorana condition.

## INTRODUCTION

Quantum walk (QW) is a simple framework to simulate quantum phenomena, and there are two main types of QW, the Continuous-Time Quantum Walk (CTQW)[1], and the Discrete-Time Quantum Walk (DTQW). In the CTQW, the walk is directly defined in the Hilbert space called the position space $\mathcal{H}_p$, and the DTQW introduces an additional space called coin space $\mathcal{H}_c$ which defines the direction of the walk. CTQW and DTQW are connected by the Szegedy Quantum Walk[2] in a certain limit[3]. Although these two types of QW have their own advantages, DTQW is much easier to implement on a near-term quantum computer[4–6].

The general DTQW was introduced as a counterpart of Random Walk[7]. This algorithm exploits quantum superposition and interference between multiple paths make it useful in a variety of applications, such as quantum search[8–11], isomorphism of graph[12–14]and quantum finance[15,16]. In fact, QW can be traced back to Feynman's checkboard. In 1965, Feynman proposed an approach of using path integral in discrete lattice in (1+1)-dimensional spacetime to derive the Dirac equation[17]. This approach inspired the investigation of algorithm called Quantum Cellular Automata (QCA)[18–20]. The QCA can be interpreted as QW algorithm related to single particle evolution[21].

As a powerful tool to simulate the Dirac fermion, the relation between DTQW and the Dirac equation has been well studied[22–24]. The relativistic quantum effect, such as Zitterbewegung and Klein's paradox, have proven to exist in the DTQW[25–27]. Therefore, the DTQW has already been used to simulate various fundamental physics, such as Anderson localization[28–30], topological phases[31,32] and neutrino oscillation[33,34]. Moreover, studies show that the DTQW can be imbedded in a curved spacetime[35,36] or coupled to electromagnetic field and another gauge field[37–40]. These results also inspire studying fundamental physics in the term of quantum information theory.

In 1928, Dirac proposed the famous Dirac equation and predicted that a fundamental particle would have an antiparticle. In 1937, Majorana hypothesized that a charged-neutral particle with a spin of ½ could be its own antiparticle, known as the Majorana fermion[41]. Compared with a Dirac fermion, a Majorana fermion must obey the Majorana condition and only has a half degree of freedom, and hence a massive Majorana fermion has two components like a Weyl fermion. Besides, for a Dirac particle one can define a fermion-number $\boldsymbol{U}(1)$ symmetry under which the Lagrangian is invariant. In contrast, for a massive Majorana fermion one cannot define such an $\boldsymbol{U}(1)$ symmetry, so a Majorana fermion does not have a fermion number. This bizarre property may make neutrino unique since it is the only known fermion which may be a Majorana one and all other known spin 1/2 fundamental particles have definite fermion numbers. Although fundamental Majorana particles have not been discovered in particle physics, experiment evidence shows a Majorana particle can emerge as a quasiparticle in condensed matter system called Majorana zero mode[42]. As mentioned previously, the topological phase in DTQW is also investigated. The result shows that the topological phase can be classified by the coin parameter, and Majorana boundary modes exist in the boundaries with different topological phases[43,44]. However, the space-time behavior and its associated dynamics of Majorana particle remains unclear. We find that DTQW with a quantum computer can achieve this goal in a rather straightforward way but with quite profound physics.

In the framework of the DTQW, not only the coin operator but also the initial state plays an important role. The initial state is specified by the nature of the particle, either a Dirac or a Majorana one, therefore DTQW provides a possibility to study in some details about spacetime behavior and associated dynamics of a Majorana particle. In this study, we follow the approach called Dirac Quantum Walk (DQW) to be described in the next section, to simulate the dynamics of a Majorana particle in (1+1)-dimensional spacetime[45] for special initial states which satisfies the Majorana condition. Our result is demonstrated by both using a simulator and a real quantum processor provided by IBM Quantum system[46]. When it comes to operation in real quantum device, we only consider the rest Majorana particle with expectation value of momentum is zero in (1+1)-dimensional with separable coin space and position space for the current limitation of quantum volume (QV) in noisy intermediate-scale quantum (NISQ) computer[47]. By using this simplified case, we analyze the dynamics of the QW with the constraint of Majorana condition.

## THEORETICAL RESULT

### Dirac Quantum Walk

In a (1+1)-dimensional spacetime, a free Dirac particle is described by the Dirac equation (with speed of light $c = 1$, and reduced Planck constant $\hbar = 1$):

$$\left(i\boldsymbol{\gamma}^{\mu}\partial_{\mu} - m\right)\Psi = 0. \tag{1}$$

Here $\Psi$ is a space and time dependent two-component spinor and $m$ is the mass of the particle. The Dirac gamma matrices, $\boldsymbol{\gamma}^{\mu}$ with $\mu = 0, 1$ satisfy the anti-commutation relation:

$$\boldsymbol{\gamma}^{\mu}\boldsymbol{\gamma}^{\nu} + \boldsymbol{\gamma}^{\nu}\boldsymbol{\gamma}^{\mu} = 2\boldsymbol{g}^{\mu\nu}\boldsymbol{I}, \tag{2}$$

where $\boldsymbol{g}^{\mu\nu} = \mathrm{diag}(1, -1)$ and $I$ is 2 by 2 identity matrix. To describe the time evolution of a Dirac particle, we rewrite the equation in the Hamiltonian form

$$i\frac{\partial}{\partial t}\Psi = \boldsymbol{H}\Psi, \tag{3}$$

where $\boldsymbol{H} = -i\boldsymbol{\gamma}^{0}\boldsymbol{\gamma}^{1}\partial_x + m\boldsymbol{\gamma}^{0}$ is the Hamiltonian operator. For this simple (1+1)-dimension spacetime case, there is no spin, and the extra degree of freedom is called chirality.

For the Weyl basis, we choose $\boldsymbol{\gamma}^{0}\boldsymbol{\gamma}^{1} = -\boldsymbol{\sigma}_z$ (this is equivalent to $\boldsymbol{\gamma}^{5}$ in 3+1 dimensions), and hence $\boldsymbol{\gamma}^{0} = \vec{\boldsymbol{n}}_{\perp z} \cdot \vec{\boldsymbol{\sigma}}$ because of the constraint described by Eq. (2), where $\vec{\boldsymbol{n}}_{\perp z} = \left[n_x, n_y, 0\right]^{\mathrm{T}}$ is an arbitrary unit vector without z-component and $\vec{\boldsymbol{\sigma}} = \left[\boldsymbol{\sigma}_x, \boldsymbol{\sigma}_y, \boldsymbol{\sigma}_z\right]^{\mathrm{T}}$ is a vector consisting of three Pauli matrices. The Hamiltonian becomes:

$$\boldsymbol{H} = i\boldsymbol{\sigma}_z\partial_x + m\vec{\boldsymbol{n}}_{\perp z} \cdot \vec{\boldsymbol{\sigma}}. \tag{4}$$

The Dirac field $\Psi$ can be projected onto the right-handed and left-handed component which are eigenvectors of $\boldsymbol{\sigma}_z$, as

$$\psi_R = \frac{\boldsymbol{I} + \boldsymbol{\sigma}_z}{2}\Psi \tag{5a}$$

and

$$\psi_L = \frac{\boldsymbol{I} - \boldsymbol{\sigma}_z}{2}\Psi, \tag{5b}$$

where $\boldsymbol{I}$ is 2 by 2 identity matrix. Therefore, the time evolution of a Dirac particle can be written as:

$$\Psi(x,t) = \begin{pmatrix} \psi_L(x,t) \\ \psi_R(x,t) \end{pmatrix} = e^{-i\boldsymbol{H}t} \begin{pmatrix} \psi_L(x,0) \\ \psi_R(x,0) \end{pmatrix} = \boldsymbol{U}(t) \begin{pmatrix} \psi_L(x,0) \\ \psi_R(x,0) \end{pmatrix}. \tag{6}$$

In a discretized spacetime, $\Delta t = \varepsilon$ is the time step and $\Delta x = c\Delta t = \varepsilon$ is the space step, and then the Dirac time evolution operator in each time step $\varepsilon$ can be approximated to the first order in $\varepsilon$, that is

$$\begin{aligned}
\boldsymbol{U}(\varepsilon) &\sim \boldsymbol{I} - i\varepsilon\boldsymbol{H} + O(\epsilon^2) \\
&= \boldsymbol{I} - i\varepsilon(i\boldsymbol{\sigma}_z\partial_x + m\vec{\boldsymbol{n}}_{\perp z}\cdot\vec{\boldsymbol{\sigma}}) + O(\epsilon^2) \\
&= (\boldsymbol{I} + \varepsilon\boldsymbol{\sigma}_z\partial_x)(\boldsymbol{I} - i\varepsilon m\vec{\boldsymbol{n}}_{\perp z}\cdot\vec{\boldsymbol{\sigma}}) + O(\epsilon^2) \\
&= \begin{pmatrix} 1+\varepsilon\partial_x & 0 \\ 0 & 1-\varepsilon\partial_x \end{pmatrix} \begin{pmatrix} 1 & -i\varepsilon m(n_x - in_y) \\ -i\varepsilon m(n_x + in_y) & 1 \end{pmatrix} + O(\epsilon^2) \\
&= \begin{pmatrix} 1+\varepsilon\partial_x & 0 \\ 0 & 1-\varepsilon\partial_x \end{pmatrix} \begin{pmatrix} 1 & -i\varepsilon m e^{-i\phi} \\ -i\varepsilon m e^{i\phi} & 1 \end{pmatrix} + O(\epsilon^2).
\end{aligned} \tag{7}$$

Here, we rewrite $n_x + in_y$ as $e^{i\phi}$ in the fourth step. When $\varepsilon$ is small enough, one can write Eq. (7) as

$$\boldsymbol{U}(\varepsilon) = \begin{pmatrix} e^{\varepsilon\partial_x} & 0 \\ 0 & e^{-\varepsilon\partial_x} \end{pmatrix} \begin{pmatrix} \cos\left(\frac{\theta}{2}\right) & -ie^{-i\phi}\sin\left(\frac{\theta}{2}\right) \\ -ie^{i\phi}\sin\left(\frac{\theta}{2}\right) & \cos\left(\frac{\theta}{2}\right) \end{pmatrix}, \tag{8}$$

where $\frac{\theta}{2} = \varepsilon m$ and $e^{\mp\varepsilon\partial_x} = (|x \pm \epsilon\rangle\langle x|)$. In Eq (8), the first term is called translation operator, and the second term is called mass operator. Consequently, the time evolution of a Dirac particle can be described by a sequence of unitary operators which is composed of rotations in spinor space and conditional space shifts, and this structure is known as the Discrete Time Quantum Walk (DTQW). Therefore, we can use DTQW algorithm the recover the Dirac time evolution.

In the DTQW algorithm, the quantum state is spanned by the coin space $\mathcal{H}_c$ and position space $\mathcal{H}_p$. In our cases, the discrete Dirac field, the right-handed and left-handed component are encoded in coin space $\mathcal{H}_c$ and the position of the particle is encoded in the position space $\mathcal{H}_p$:

$$\Psi(x,t) = (\psi_L(x,t)|0\rangle_c + \psi_R(x,t)|1\rangle_c) \otimes |x\rangle \tag{9}$$

where $|0\rangle_c$ is $[1,0]^{\mathrm{T}}$, $|1\rangle_c$ is $[0,1]^{\mathrm{T}}$ and $x, t \in \mathbb{Z}$ which label the discrete spacetime coordinate. Therefore, the state of the particle after time step $t$ is

$$\Psi(t) = \sum_{x\in\mathbb{Z}} \begin{pmatrix} \psi_L(x,t) \\ \psi_R(x,t) \end{pmatrix} = \boldsymbol{W}^t \sum_{x\in\mathbb{Z}} \begin{pmatrix} \psi_L(x,0) \\ \psi_R(x,0) \end{pmatrix}, \tag{10}$$

where

$$\boldsymbol{W} = \boldsymbol{S}(\boldsymbol{B} \otimes \boldsymbol{I}_p). \tag{11}$$

In Eq. (11), the coin operator $\boldsymbol{B}$ is an analog of the mass operator in Eq. (8) and $\boldsymbol{I}_p = \sum_{x\in\mathbb{Z}}|x\rangle\langle x|$, and hence the mass operator indeed acts on the all position ubiquitously. For general DTQW, the coin operator is any unitary matrix, however, the our our case, the coin operator is constrained by Eq. (2) with Dirac dynamics. However, the Ref.[25] shows that even without this constraint the bizarre phenomenon of Dirac dynamics such as Zitterbewegung and Klein paradox still occurred in DTQW. On the other hand, the conditional shift operator $\boldsymbol{S}$ in Eq. (11) is an analog of translation operator and the shift direction is controlled by the state of coin space $\mathcal{H}_c$, but the conditional shift operator $\boldsymbol{S}$ acts on the all position ubiquitously:

$$\boldsymbol{S} = |0\rangle_{c\,c}\langle 0| \otimes \sum_{x\in\mathbb{Z}} |x-1\rangle\langle x| + |1\rangle_{c\,c}\langle 1| \otimes \sum_{x\in\mathbb{Z}} |x+1\rangle\langle x|. \tag{12}$$

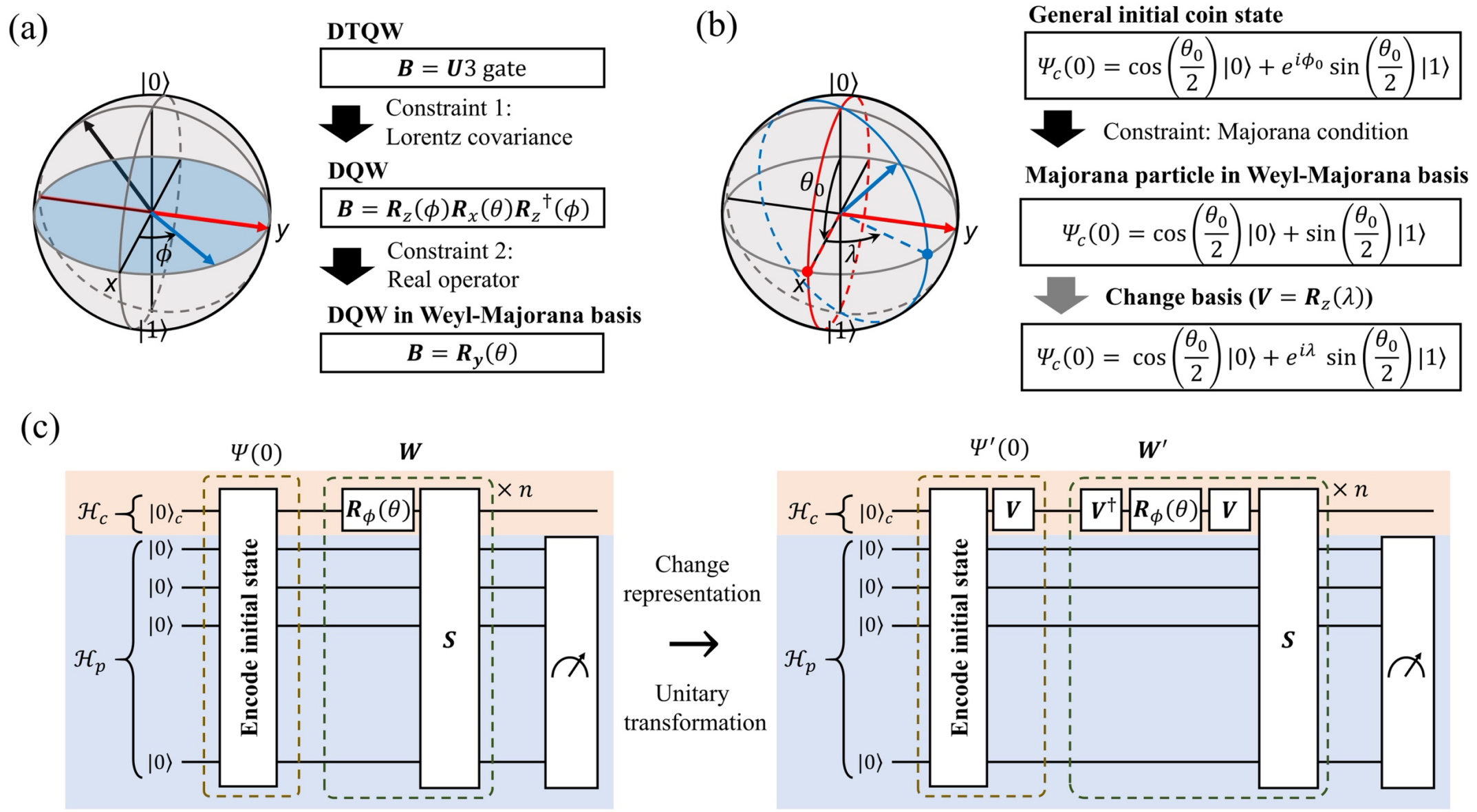

**Fig. 1 Methodology of simulating dynamics of Majorana particle (a)** We imposed two constraints in DTQW. The first one is Lorentz covariance, and this constraint projects the rotation axis of the coin onto the *x*-*y* plane ($0 \le \phi < 2\pi$). The second one is a real operator (Weyl-Majorana basis), and this constraint projects the rotation axis of the coin to the *y*-axis ($\phi = \pm\frac{\pi}{2}$). **(b)** Geometric relation between the Majorana condition and basis choice in DQW. The Majorana condition constrain the spinor located in the biggest circle that is perpendicular to the rotation axis. **(c)** DQW algorithm goes through the change of the basis. It encodes the initial state to the quantum registers at first, and then applies the walk operator to simulate the time evolution. When the basis changes, applying the corresponding unitary transformation $\boldsymbol{V}$.

Now, we define the DTQW with the Lorentz invariance constraint for the rotation axis in the *x*-*y* plane of $\vec{\boldsymbol{\sigma}}$ space as Dirac Quantum Walk (DQW) as shown in **Fig. 1(a)**. According to Eq. (8), the coin operator of DQW is parameterized by the parameter $\phi$ which corresponds to the freedom of basis choice and the parameter $\theta$ is proportional to the mass of the particle, that is

$$\boldsymbol{B}_\phi(\theta) = \begin{pmatrix} \cos\left(\frac{\theta}{2}\right) & -ie^{-i\phi}\sin\left(\frac{\theta}{2}\right) \\ -ie^{i\phi}\sin\left(\frac{\theta}{2}\right) & \cos\left(\frac{\theta}{2}\right) \end{pmatrix}$$
$$= \boldsymbol{R}_z(\phi)\boldsymbol{R}_x(\theta)\boldsymbol{R}_z{}^\dagger(\phi), \tag{13}$$

where

$$\boldsymbol{R}_x(\theta) = \begin{pmatrix} \cos\left(\frac{\theta}{2}\right) & -i\sin\left(\frac{\theta}{2}\right) \\ -i\sin\left(\frac{\theta}{2}\right) & \cos\left(\frac{\theta}{2}\right) \end{pmatrix}, \tag{14}$$

and

$$\boldsymbol{R}_z(\phi) = e^{-i\frac{\phi}{2}\sigma_z}. \tag{15}$$

In the following discussion, we define two special common bases: one with $\phi = 0$ , the Weyl-Dirac basis ($\boldsymbol{\gamma}^0 = \boldsymbol{\sigma}_x, \boldsymbol{\gamma}^1 = i\boldsymbol{\sigma}_y$), and the other with $\phi = \frac{\pi}{2}$ , the Weyl-Majorana basis ($\boldsymbol{\gamma}^0 = \boldsymbol{\sigma}_y, \boldsymbol{\gamma}^1 = -i\boldsymbol{\sigma}_x$). In the Weyl-Dirac basis, the coin operator $\boldsymbol{B}_D(\theta)$ rotate internal degree of freedom around the *x*-axis**:**

$$\boldsymbol{B}_D(\theta) = \boldsymbol{R}_x(\theta) \tag{16}$$

This symmetrical coin operator is commonly used, for example, in Ref.[20,22,24,26,27]. On the other hand, in the Weyl-Majorana basis, the coin operator $\boldsymbol{B}_M(\theta)$ rotate internal degree of freedom around the *y*-axis as shown in **Fig. 1 (a)**, and is a real matrix. This coin operator is used in the Ref.[31,32] to analysis the topology of QW.

$$\boldsymbol{B}_M(\theta) = \boldsymbol{R}_y(\theta) = \begin{pmatrix} \cos\left(\frac{\theta}{2}\right) & -\sin\left(\frac{\theta}{2}\right) \\ \sin\left(\frac{\theta}{2}\right) & \cos\left(\frac{\theta}{2}\right) \end{pmatrix} \tag{17}$$

In this case, the gamma matrices satisfy $(i\boldsymbol{\gamma}^\mu)^* = i\boldsymbol{\gamma}^\mu$, i.e., $\boldsymbol{\gamma}^\mu$'s are pure imaginary, where $*$ denotes complex-conjugation, hence the time revolution operator is real.

**Majorana particle in DQW**

A Majorana particle is its own antiparticle. Specifically, the particle satisfies the constraint which the wave function of the particle $\Psi$ is identical to its charge conjugate wave function $\widehat{\Psi}$, that is

$$\Psi = \widehat{\Psi}. \tag{18}$$

This relation is referred as the Majorana condition. In Weyl-Majorana basis, $\widehat{\Psi} = \Psi^*$ and thus $\Psi$ is real. This implies that when the initial state of the QW is real function, and applying the walk operator $\boldsymbol{W}$ does not change this property, because all the components of $\boldsymbol{W}$ are real in the Weyl-Majorana basis.

From Eq. (9), the general initial state of Dirac field can be encoded as

$$\Psi(0) = \sum_{x\in\mathbb{Z}} \left( \cos\left(\frac{\theta_0(x)}{2}\right) |0\rangle_c + e^{i\phi_0} \sin\left(\frac{\theta_0(x)}{2}\right) |1\rangle_c \right) \otimes D(x)|x\rangle. \tag{19}$$

In the coin space, the $\cos\left(\frac{\theta_0(x)}{2}\right)$ represent left-handed component and $\sin\left(\frac{\theta_0(x)}{2}\right)$ represent right-handed component of the field and the $\phi_0$ is the relative phase between these two components. We have inserted the complex number $D(x)$ space distribution representing a realistic wave packet which describes the component of field in position $x$ and satisfies that $\sum_{x\in\mathbb{Z}} |D(x)|^2 = 1$.

Despite of the fact that the initial state as a wave packet can represent phycal significance of a particle[20,22], we use the general initial state of QW given by

$$\Psi(0) = \left( \cos\left(\frac{\theta_0}{2}\right) |0\rangle_c + e^{i\phi_0} \sin\left(\frac{\theta_0}{2}\right) |1\rangle_c \right) \otimes |x = 0\rangle \tag{20}$$

to simulate the dynamic behavior of a particle for the sake of reducing quantum gate circuit depth without loss of generality. Eq. (20) describes a particle at $x = 0$ with expectation value of initial momentum is zero. Notice that the coin space and the position space are separable, but it is not the case in Eq. (19).

In Weyl-Majorana basis, the charge conjugation of Eq. (20) is

$$\widehat{\Psi}(0) = \left( \cos\left(\frac{\theta_0}{2}\right) |0\rangle_c + e^{-i\phi_0} \sin\left(\frac{\theta_0}{2}\right) |1\rangle_c \right) \otimes |x = 0\rangle, \tag{21}$$

and when $\phi_0 = 0$, $\Psi = \widehat{\Psi}$ and the component of $\Psi(0)$ is pure real, hence it satisfies the Majorana condition.

One may choose a different basis of the gamma matrices $(\boldsymbol{\gamma}')^\mu$, and the two sets of gamma matrices are related by a unitary transformation $\boldsymbol{V}$, that is

$$(\boldsymbol{\gamma}')^\mu = \boldsymbol{V}\boldsymbol{\gamma}^\mu\boldsymbol{V}^\dagger. \tag{22}$$

The walk operator and state must undergo the same unitary transformation, and thus we obtain

$$\boldsymbol{W}' = (\boldsymbol{V} \otimes \boldsymbol{I}_p)\boldsymbol{W}(\boldsymbol{V}^\dagger \otimes \boldsymbol{I}_p) \tag{23}$$

and

$$\Psi' = (\boldsymbol{V} \otimes \boldsymbol{I}_p)\Psi. \tag{24}$$

This leads to

$$\Psi'^{\dagger}\boldsymbol{W}'\Psi' = \Psi^{\dagger}\boldsymbol{W}\Psi. \tag{25}$$

Therefore, Eq. (25) shows that expectation value of DQW is independent of the basis choice, i.e., physical results do not depend on the basis.

In the Weyl basis, one can notice that, the unitary operator $\boldsymbol{V} = \boldsymbol{R}_z(\lambda)$, where $\lambda$ is the angle between the rotation axes in $\boldsymbol{W}$ and $\boldsymbol{W}'$. In addition, using Eq. (24), we can find that the Majorana condition in different basis becomes

$$(\boldsymbol{V}^{\dagger} \otimes \boldsymbol{I}_p)\Psi' = \left((\boldsymbol{V}^{\dagger} \otimes \boldsymbol{I}_p)\Psi'\right)^{*} \tag{26}$$

or

$$\Psi' = (\boldsymbol{V} \otimes \boldsymbol{I}_p)(\boldsymbol{V}^{\boldsymbol{T}} \otimes \boldsymbol{I}_p)\Psi'^{*} = (e^{-i\lambda\sigma_z} \otimes \boldsymbol{I}_p)\Psi'^{*}. \tag{27}$$

This implies that they are Weyl-Majorana particles, because charge conjugate commutes with $\boldsymbol{\sigma_z}$, i.e. the charge conjugate of field $\psi_R$ ($\psi_L$) are also right-handed (left-handed). But this is not the case in 3+1-dimensional spacetime[48].

The geometric relation between the Majorana condition and basis choice of DQW is shown in **Fig. 1 (b)**. The Majorana particle in all bases is the spinor located in the biggest circle that is perpendicular to the corresponding rotation axis. Hence it only has a half degree of freedom compared to the Dirac particle. For instance, in Weyl-Dirac basis ($\phi = 0$), the Majorana condition is $\Psi = (\boldsymbol{\sigma}_z \otimes \boldsymbol{I}_p)\Psi^{*}$. In this case, $\phi_0 = -\frac{\pi}{2}$, the component of $|1\rangle$ is pure imaginary, and hence satisfies the Majorana condition.

The quantum circuit of DQW under the basis change is shown in **Fig. 1(c)**. The unitary transformation only changes the coin operator and the initial state, because the control shift operator is invariant under the unitary transformation namely $(\boldsymbol{V} \otimes \boldsymbol{I}_p)\boldsymbol{S}(\boldsymbol{V}^{\dagger} \otimes \boldsymbol{I}_p) = \boldsymbol{S}$.

**Second order correction**

For general DTQW, as the counterpart of Random Walk, we apply $\boldsymbol{B}_{\phi}(\theta)$ at first, because it is analog of coin tossing. However, $\left(\boldsymbol{B}_{\phi}(\theta) \otimes \boldsymbol{I}_p\right)$ and $\boldsymbol{S}$ do not commute, and hence the above approximation is only accurate up in $\varepsilon$ to first order. Considering the DQW with $\boldsymbol{B}_{\phi}(\theta)$ and $\boldsymbol{S}$ reversed:

$$\Psi(t) = \boldsymbol{W}_{BS}^{t}\Psi(0) \tag{28}$$

$$= \left(\boldsymbol{B}_{\phi}(\theta) \otimes \boldsymbol{I}_p\right) \boldsymbol{W}_{SB}^{t}\left(\boldsymbol{B}_{\phi}^{-1}(\theta) \otimes \boldsymbol{I}_p\right) \Psi(0) \tag{29}$$

where $\boldsymbol{W}_{SB} = \boldsymbol{S}\left(\boldsymbol{B}_{\phi}(\theta) \otimes \boldsymbol{I}_p\right)$ and $\boldsymbol{W}_{BS} = \left(\boldsymbol{B}_{\phi}(\theta) \otimes \boldsymbol{I}_p\right)\boldsymbol{S}$, the final position of initial state $\Psi(0)$ is equivalent to the initial state $\left(\boldsymbol{B}_{\phi}^{-1}(\theta) \otimes \boldsymbol{I}_p\right)\Psi(0)$ in original DQW, because the last coin operator in Eq. (29) do not affect the particle position. Obviously, when the $\theta$ is small enough, the two walk operators will get the close position distribution. This show the error of the simulate the Dirac evolution is associated with mass term.

To achieve a higher order approximation in $\varepsilon$, one needs to make modification to how one approximate $\boldsymbol{W} = e^{-i\boldsymbol{H}\varepsilon}$ to combinations of the coin operator $\boldsymbol{B}$ and conditional shift operator $\boldsymbol{S}$. It has been proved that the higher order approximation can be achieve by the symmetric decomposition (see appendix in Ref.[49]), and we find that there are two solutions to the second order in $\varepsilon$. One is given by

$$\boldsymbol{W}_{BSB} = \boldsymbol{B}_{\phi}\left(\frac{\theta}{2}\right)\boldsymbol{S}\boldsymbol{B}_{\phi}\left(\frac{\theta}{2}\right) \tag{30}$$

and there other is

$$\boldsymbol{W}_{SBS} = \boldsymbol{S}_{+}\boldsymbol{B}_{\phi}(\theta)\boldsymbol{S}_{-} \tag{31}$$

where

$$S_- = \begin{pmatrix} T_- & 0 \\ 0 & I \end{pmatrix}, \qquad S_+ = \begin{pmatrix} I & 0 \\ 0 & T_+ \end{pmatrix} \tag{32}$$

For $W_{BSB}$, $S$ sandwiched between the square root of $B_\phi(\theta)$ ($\sqrt{B_\phi(\theta)} = B_\phi\left(\frac{\theta}{2}\right)$). Once again, one can find that the final position probability of initial state $\Psi(0)$ after applying $W_{BSB}$ is equivalent to $B_\phi\left(-\frac{\theta}{2}\right)\Psi(0)$ in original DQW, and hence this modified QW operator can present the probability distribution between the $W_{SB}$ and $W_{BS}$. It also has the advantage of easily to be realized in quantum circuit implementation and compatible with existing efficient quantum walk algorithm[50,51] with same complexity. In real quantum device, this higher order correction can get a much precise outcome with the same quantum volume and it is good for QEC for NISQ computer.

On the other hand, for $W_{SBS}$, we split $S$ into two part, $S_+$ and $S_-$, because $S = S_+S_- = S_-S_+$. This modified DQW is equivalent to Dirac Cellular Automata (DCA)[13,14] but with phase $\phi$ in off-diagonal term (the freedom to change the basis). The Dirac Cellular Automata is a discrete spacetime model with casual unitary evolution that derived from the assumption of homogeneity, parity and time reversal invariance.

As mentioned in introduction, the DTQW is known as a subset of QCA. In the language of Cellular Automata, in DTQW, the state of next iteration in position $x$ is dependent only on the state of its nearest neighboring. Nevertheless, in QCA, the state of next iteration in position $x$ not only corresponding to its neighboring, but also itself.

The relation between DCA, DQW and Dirac Equation is bridged from Split-step DTQW (SQW)[15]. Briefly, in SQW, each step of DTQW is split into two half-steps:

$$W_{SQW} = S_+(B_2 \otimes I)S_-(B_1 \otimes I), \tag{33}$$

When the first coin operator $B_1$ is Identity matrix, the SQW becomes DCA. Oppositely, when the second coin operator $B_2$ is Identity matrix, the SQW becomes DQW.

**Entanglement generated by DTQW**

To understand the property of the Majorana particle from perspective of quantum information, it is worthy to analyze the entanglement entropy generated by the time evolution operators under the Majorana condition. Here we define the entanglement by the von Neumann entropy, and the density matrix of the initial state in Eq. (20) is given by

$$\rho(0) = \begin{pmatrix} \cos^2\left(\frac{\theta_0}{2}\right) & \cos\left(\frac{\theta_0}{2}\right)\sin\left(\frac{\theta_0}{2}\right)e^{-i\phi_0} \\ \cos\left(\frac{\theta_0}{2}\right)\sin\left(\frac{\theta_0}{2}\right)e^{i\phi_0} & \sin^2\left(\frac{\theta_0}{2}\right) \end{pmatrix} \otimes |x=0\rangle\langle x=0|. \tag{34}$$

Form Eq. (10), the density matrix after $t$ steps is

$$\rho(t) = W^t\rho(0)(W^\dagger)^t. \tag{35}$$

Then the entanglement entropy between the coin space and position space at step $t$ is given by

$$S(\rho_c(t)) = -\mathrm{Tr}_c(\rho_c(t)\log(\rho_c(t))) = S(\rho_x(t)), \tag{36}$$

where $\rho_c(t) = \mathrm{Tr}_x(\rho(t))$ and $\rho_x(t) = \mathrm{Tr}_c(\rho(t))$ are reduced density matrices which take partial trace respect to position space $\mathcal{H}_p$ and coin space $\mathcal{H}_c$, respectively. In Ref.[27], the results shown that the values of entanglement generated by DTQW significantly depend on the initial state. Consider three different initial state with $\phi_0 = 0, \frac{\pi}{2}, \frac{\pi}{4}$ with $\theta_0$ fixed in $\frac{\pi}{2}$. From their study, the entanglement generated by the $W_{SB}$ in these three different initial states is fluctuated around same mean values (see Fig. 2(a)). Moreover, comparing their result of $W_{SBS}$ with the entanglement generated by $W_{BSB}$, we find that both the second order approximation operator reduce the fluctuation of the entanglement and make the Majorana particle entanglement stronger than the Dirac one (see **Fig.2 (b)** and **(c)**).

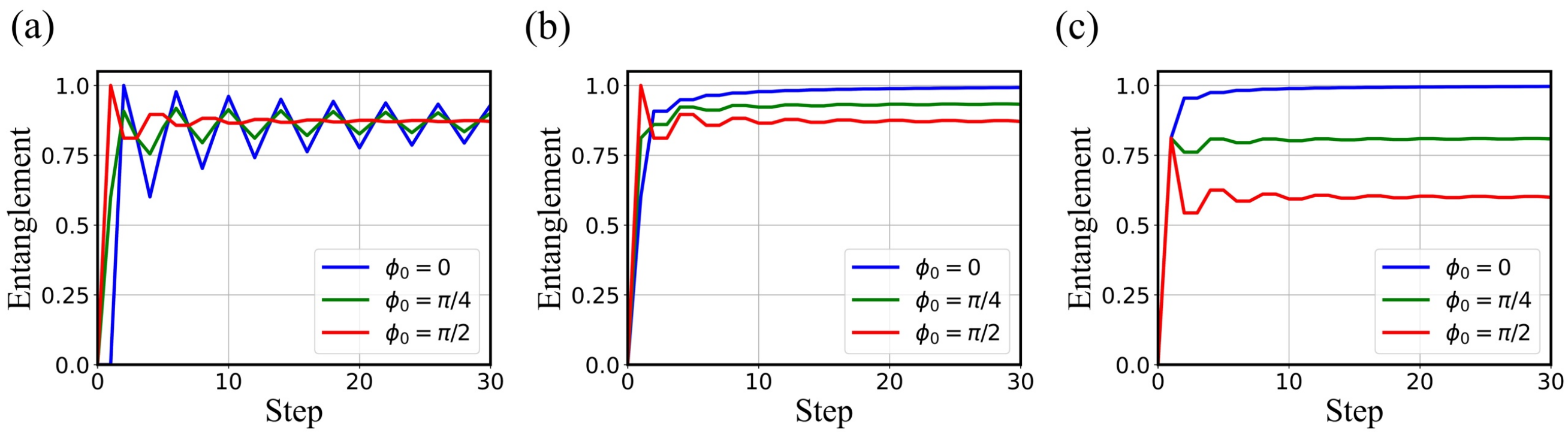

**Fig. 2 The entanglement between the coin space and position space with different time evolution operator in Weyl-Majorana basis.** The parameter $\theta_0$ of initial state is fixed in $\frac{\pi}{2}$, and the state satisfied Majorana condition when $\phi_0 = 0$. (a) $\boldsymbol{W}_{SB}$, (c) $\boldsymbol{W}_{BSB}$, and (d) $\boldsymbol{W}_{SBS}$.

**Numerical simulations**

In a DTQW algorithm, the readout of the quantum state is shown as the position distribution, and the position distribution describe the probability of finding a particle in the position $x$ at time $t$, that is

$$P(x,t) = |\psi_L(x,t)|^2 + |\psi_R(x,t)|^2. \tag{37}$$

In other words, we only measure the qubits state of position space.

For the numerical simulations, we consider an initial state as Dirac plane wave solution with $p = 0$ in Weyl-Majorana basis (see Ref.[45] for detail) at first, and encode this state in quantum registers as

$$\Psi_D(0) = \frac{1}{\sqrt{2}}\binom{1}{i} \otimes \frac{1}{\sqrt{N}}\sum_{x=0}^{N-1}|x\rangle, \tag{38}$$

where $k$ is number of qubits of position space. This uniform position distribution (or uniform superposition) can be easily created by applying the Hadamard gate to all the qubits of position space. To construct the Majorana plane wave with $p = 0$, we consider the general solution of Dirac particle $\Psi(0) = c_0^+\Psi_D(0) + c_0^-\Psi_D^*(0)$, where $c_0^+$ and $c_0^-$ are arbitrary complex number and $\Psi_D^*(0)$ is an eigenstate with negative energy. If we want to describe a Majorana plane wave with $p = 0$, then its wavefunction $\Psi$ must satisfy the Majorana condition $\Psi = \Psi^*$. This implies that $c_0^+ = (c_0^-)^*$ and thus $|c_0^+| = \frac{1}{\sqrt{2}}$. Here, we define $c_0^+ = \frac{1}{\sqrt{2}}e^{-i\delta}$, and then obtain

$$\Psi_M(0) = \frac{1}{2}\begin{pmatrix} e^{-i\delta} + e^{+i\delta} \\ ie^{-i\delta} - ie^{+i\delta}\end{pmatrix} \otimes \frac{1}{\sqrt{N}}\sum_{x=0}^{N-1}|x\rangle$$

$$= \begin{pmatrix}\cos(\delta) \\ \sin(\delta)\end{pmatrix} \otimes \frac{1}{\sqrt{N}}\sum_{x=0}^{N-1}|x\rangle \tag{39}$$

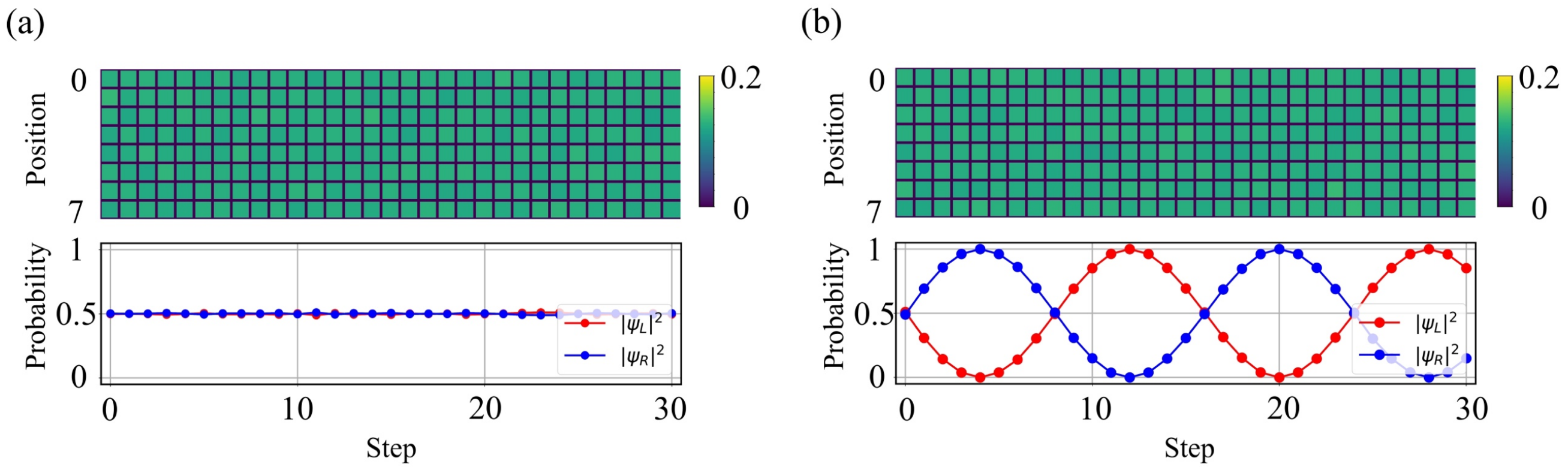

**Fig. 3 The time evolution of Dirac plane wave and Majorana plane wave (with $\delta = \frac{\pi}{4}$) with $p = 0$ and $\theta = \frac{\pi}{8}$ in Weyl-Majorana basis. The upper part of the picture demonstrates the position distribution which is encode by 3 qubits. The lower part of the picture shows the internal degree of the freedom of the particle.** (a) Dirac plane wave is a stationary state, because it is the eigenstate of Dirac equation with eigenvalue $E = E_{p=0} = m$. (b) Although position distribution of the Majorana plane wave is stationary, its internal degree of freedom oscillation with the frequency $\omega = m \equiv \frac{\theta}{2}$.

After $t$ steps of DQW, we obtain

$$\Psi_D(t) = \boldsymbol{W}^t \Psi_D(0) = \frac{1}{\sqrt{2}} \begin{pmatrix} 1 \\ i \end{pmatrix}_{\mathrm{c}} e^{-i\omega t} \otimes \frac{1}{\sqrt{N}} \sum_{x=0}^{N-1} |x\rangle\,, \tag{40}$$

and

$$\Psi_M(t) = \boldsymbol{W}^t \Psi_M(0) = \begin{pmatrix} \cos(\omega t + \delta) \\ \sin(\omega t + \delta) \end{pmatrix} \otimes \frac{1}{\sqrt{N}} \sum_{x=0}^{N-1} |x\rangle\,, \tag{41}$$

where $\omega = m \equiv \frac{\theta}{2}$ is the oscillation frequency. For $\Psi_D$, the walk operator $\boldsymbol{W}$ do nothing but add a global phase $-\omega t$ to the state, because it is the eigenstate of Dirac equation with eigenvalue $E = E_{p=0} = m$ (see **Fig.3 (a)**). Moreover, one can find that, walk operator add a global phase $\omega t$ for antiparticle $\Psi_D^*$, because it is the eigenstate of Dirac equation with negative energy $E = -E_{p=0} = -m$. However, for $\Psi_M$, we find that upper and lower component probabilities are $\cos^2(\omega t + \delta)$ and $\sin^2(\omega t + \delta)$, respectively. The phenomenon is called Majorana oscillation[52] (see **Fig.3 (b)**). Therefore, in Majorana case, the probability for the state in upper or low component is sensitive to the particle mass.

In **Fig. 4**, we demonstrate the time evolution of the initial state defined in Eq. (20), and use $\Psi_M(0) = \frac{1}{\sqrt{2}}(|0\rangle + |1\rangle)_c \otimes |x = 0\rangle$ to represent Majorana particle and $\Psi_D(0) = \frac{1}{\sqrt{2}}(|0\rangle + i|1\rangle)_c \otimes |x = 0\rangle$ to represent Dirac particle. These particles are propagated by the walk operator (a) $\boldsymbol{W}_{SB}$, (b) $\boldsymbol{W}_{SB}$, (c) $\boldsymbol{W}_{BSB}$, and (d) $\boldsymbol{W}_{SBS}$ in Weyl-Majorana basis. We varied $\theta$ to simulate the propagation of the particle with mass (i) $\theta = 0$, (ii) $\theta = \frac{\pi}{2}$ and (iii) $\theta = \pi$. The $\theta$ controls the propagation of the field, and the dispersion relation[22,26] of the field is

$$\cos\big(\omega(k)\big) = \cos\left(\frac{\theta}{2}\right)\cos(k)\,, \tag{42}$$

where $\omega$ is the frequency and $k$ is the wavenumber of the field. As a result, the particle propagates in the maximum group velocity $v = \cos\left(\frac{\theta}{2}\right)$, and hence the position distribution under after $t$ steps is spread over the interval $\left[-t\cos\left(\frac{\theta}{2}\right), t\cos\left(\frac{\theta}{2}\right)\right]$. The spacetime diagram with different walk operator shows that the massless particle ($\theta = 0$) which travels at the speed of light ($v = c = 1$). In this case, the coin operator is an identity matrix, and thus $\psi_L$and $\psi_R$ are decoupled. When $\theta = \frac{\pi}{2}$, such a coin operator is called balance coin, because the particle has the same probability to go either right or left in each position of the spacetime. From the path integral perspective, all possible paths interfere with one another, and we get the result that a massive particle spreads slower than a massless particle. Moreover, when $\theta = \pi$, the massive particle has reached the upper limit of mass in this system, we find that the propagation speed becomes zero ($v = 0$). In this case, the particle changes the direction at each step, because the coin operator exchange $\psi_L$ and $\psi_R$, and therefore causes the particle stayed in the original position.

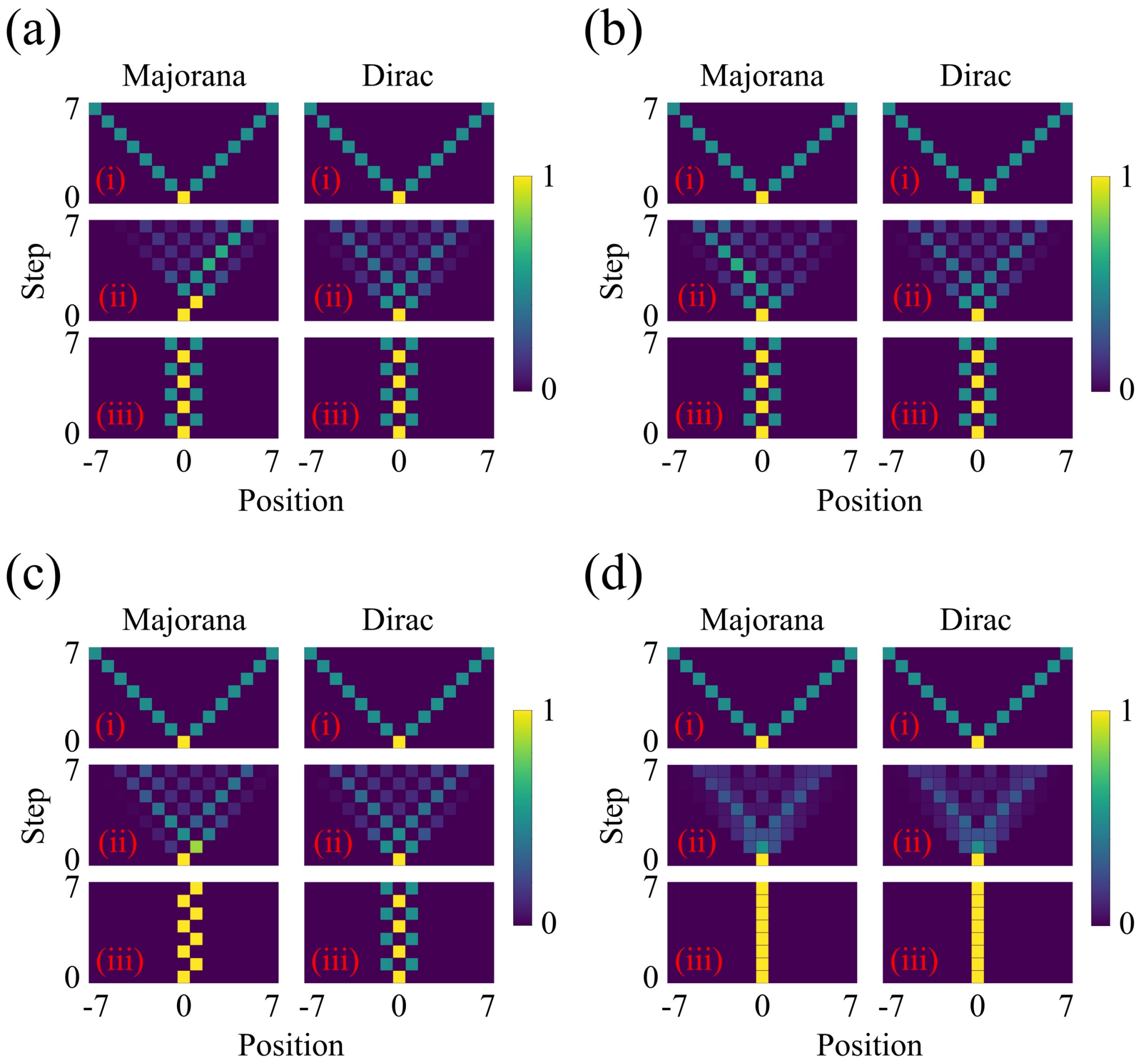

**Fig. 4 The time evolution of Majorana particle and Dirac particle in Weyl-Majorana basis.** Time evolution under the operator (a) $\boldsymbol{W}_{SB}$, (b) $\boldsymbol{W}_{BS}$, (c) $\boldsymbol{W}_{BSB}$, and (d) $\boldsymbol{W}_{SBS}$ for different mass (i) $\theta = 0$, (ii) $\theta = \frac{\pi}{2}$ and (iii) $\theta = \pi$.

However, the mass not only limit the propagation speed of the particle, but also affect the position distribution under different walk operators. When the $\theta = 0$, all the walk operators are identical. Therefore, the position distributions of the massless particle are the same. On the other hand, when the mass of particle increased, as shown **Fig. 4(a), (b)** and **(c),** the position distribution of Majorana became different, but it is not the case for Dirac particle. This is because the state of coin space in Dirac particle is eigenvector of $\boldsymbol{\sigma}_y$, in other words, coin state of the particle is unchanged under the action of coin operator. But, in **Fig. 4(d),** the result shows that the Majorana particle and the Dirac particle has same position distribution. The higher order correction $\boldsymbol{W}_{BSB}$ and $\boldsymbol{W}_{SBS}$ proposed in this article indeed provide a better numerical outcomes than usual first order correction in DTQW algorithm.

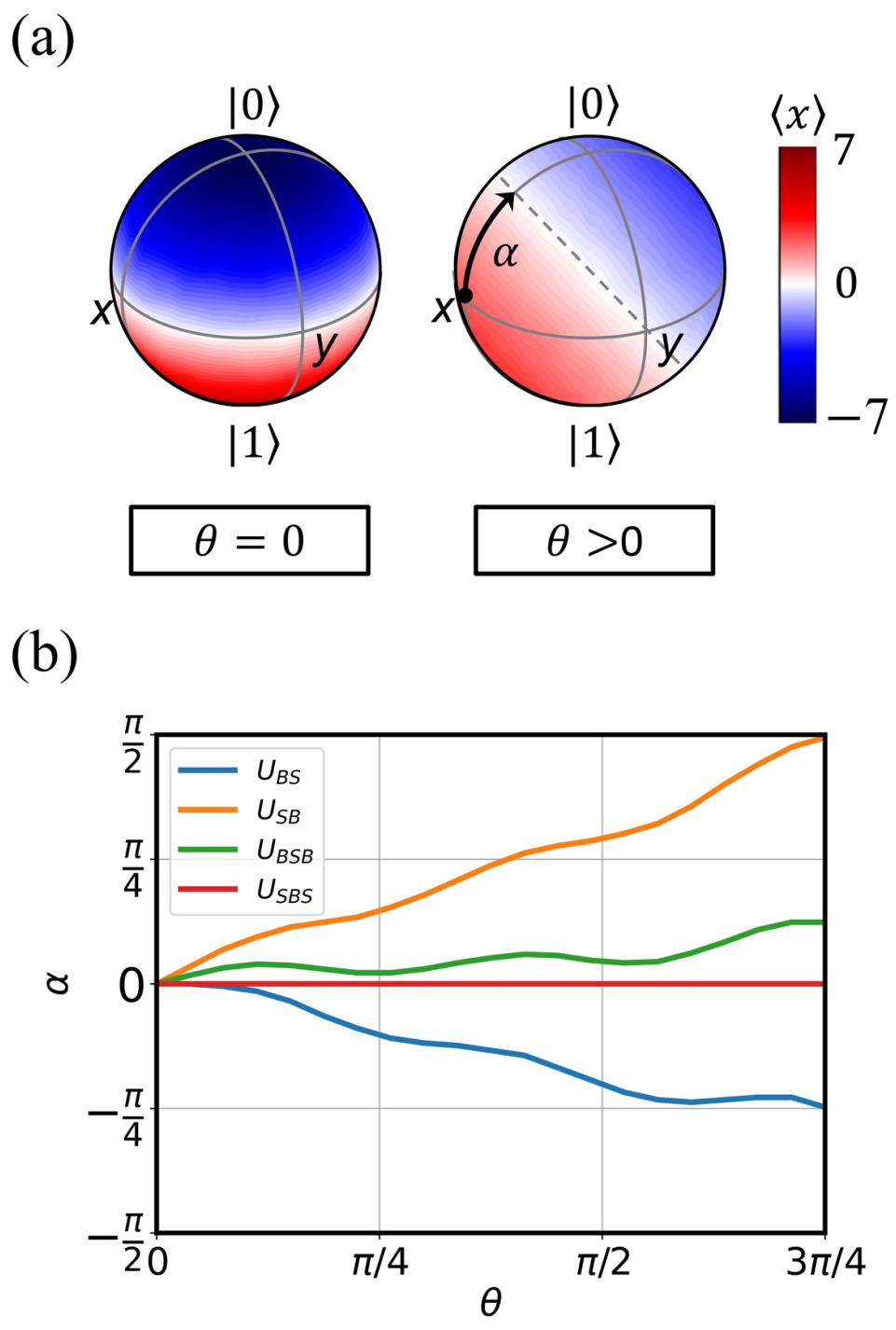

**Fig. 5 The expectation value of $x$ after 7 steps in Weyl-Majorana basis.** (a) The effect of mass term on the probability distribution is present in the Bloch sphere. (b) The shift angle $\alpha$ labeled in (a) for different walk operators.

To further investigate the effect of mass term in the QW, we simulated all possible initial states in the Bloch sphere, and then evaluate the expectation value of position $x$:

$$\langle x \rangle = \sum_x x \cdot P(x)\,, \tag{43}$$

where $P(x)$ is the probability to find the particle in the position $x$. As shown in **Fig. 5(a)**, the massless particles have symmetric distribution when their initial state is in the equator of the Bloch sphere, and we notice that mass term will rotates the sphere around y-axis. As a result, the Majorana condition cause the mass term has a stronger influence on the Majorana particle than on the Dirac particle. In addition, one can find that the distribution is symmetric under the transformation $\phi_0 \to -\phi_0$, and it means that a particle and its antiparticle have the same position distribution. That is because dispersion relation shows that particle and its antiparticle have the same velocity. In addition, for the Dirac particle, there are two states preserve the symmetry under the variation of the mass, that is $\Psi_D$ and $\widehat{\Psi}_D$, because they are the eigenvectors of the coin operator. **Fig. 5(b)** demonstrate the rotation angle $\alpha$ under different walk operators, it shows the second correction reduce the rotation angle $\alpha$, and $\alpha = 0$ for $\boldsymbol{W}_{SBS}$. This result shows that the high order correction is important for simulation of a Majorana particle.

**Experimental realization**

For near-term quantum computer, to apply same strategy to reduce the depth of the DTQW like Ref.[51,53,54], we can easily simulate the Dynamic of Majorana particle in current existing quantum computer. However, when it comes to recover the Dirac equation with second order approximation in the near-term quantum computer, for example, the DCA algorithm needs lots more quantum gate-depth than the DTQW, because it has to considers the state of the original cell at each step. As a result, the best way to achieve second order approximation is only with $\boldsymbol{W}_{BSB}$, and the second order approximation can still be easily achieved by change the rotation angle of the first coin operator from $\theta$ to $\frac{\theta}{2}$, and hence it is compatible with the efficient DTQW algorithm proposed in Ref[51]. In **Fig. 6 (a)**, the spacetime diagram closely matched the ideal distribution for the first few steps, and then became more dissimilar as the steps increase. The noises and errors let the possible routes can be developed into a more complicated maps instead of straight boundary lines in the spacetime diagrams. Meanwhile, when we simulate in Weyl-Dirac basis (see **Fig. 6 (b)**), the result shows that probability distribution is independent of the representation, and thus consistent with Eq. (25). Unfortunately, we cannot use available IBM Q to achieve second order approximation of $\boldsymbol{W}_{SBS}$ as we did in previous section from numerical simulation for the large errors and noises induced from more quantum gate depths.

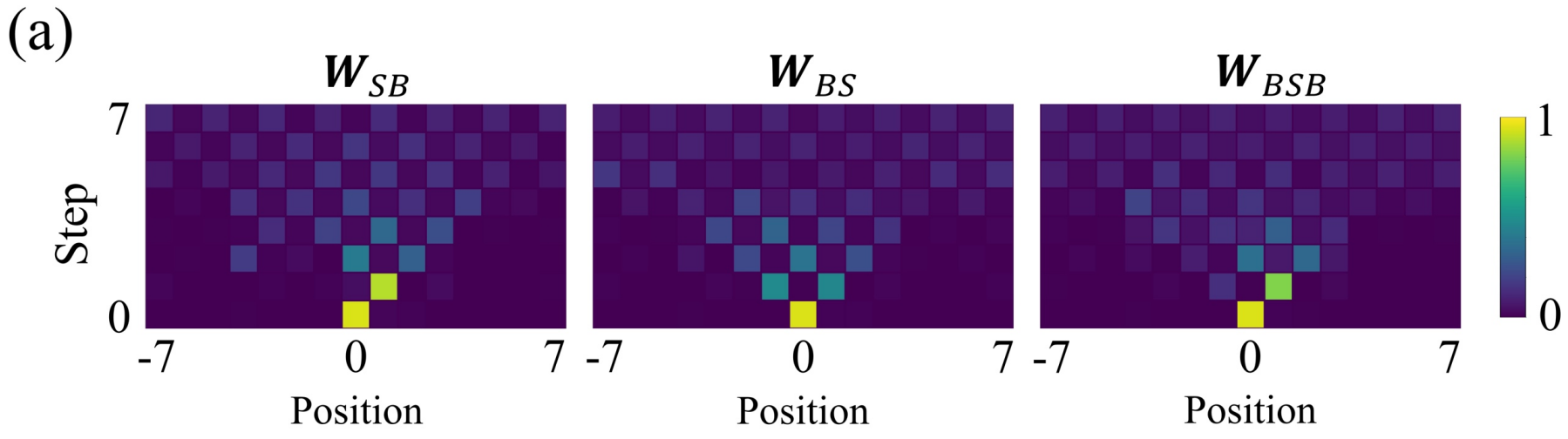

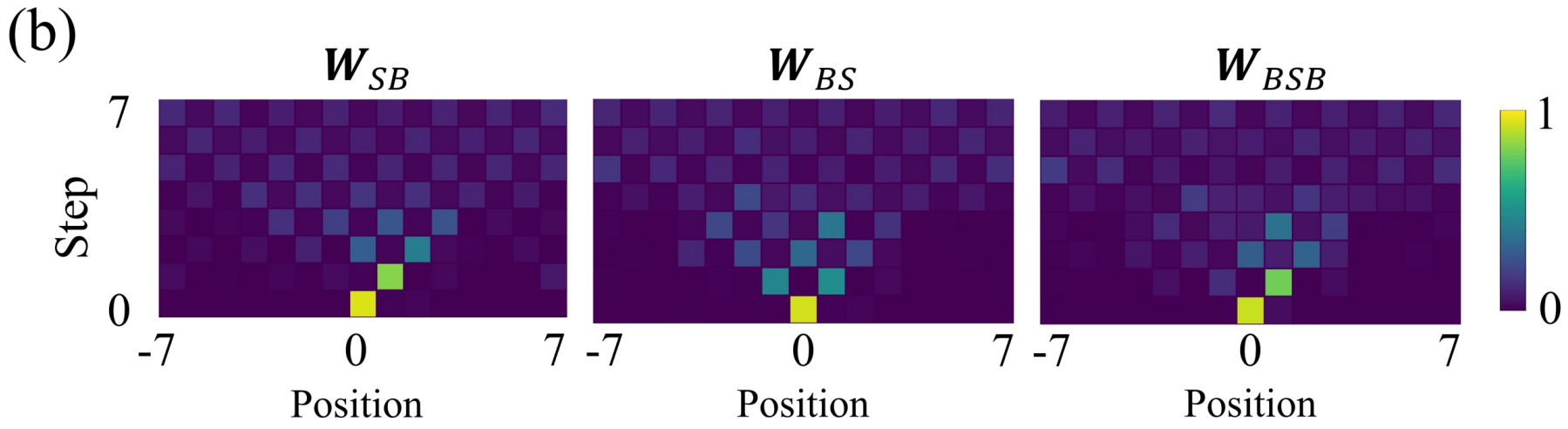

**Fig. 6 Real device simulation of time evolution of a Majorana particle via different walk operator in (a)** Weyl-Majorana basis and **(b)** Weyl-Dirac basis. The initial state for the Majorana particle is $\Psi(0) = \frac{1}{\sqrt{2}}(|0\rangle + |1\rangle)_c \otimes |x = 0\rangle$, and the $\theta = \frac{\pi}{2}$. The corresponding ideal time evolution is shown in **Fig. 3**.

## DISCUSSIONS

We propose a DTQW algorithm with constrains both on the coin operator and initial state to simulate the Majorana dynamics. It was being found that DTQW with the coin operator satisfying the Lorentz covariance can be use to describe a Dirac particle. In this work, we imposed another constraint, the Majorana condition, in the general initial state of QW (see Eq. (20)), and demonstrate the geometric relation between the Majorana condition and the basis choice in DQW. Using both the numerical simulator and real quantum device provided by IBM Quantum system, we simulate the dispersion of Majorana particles and Dirac particles in a one-dimension space. We find that the mass affects the probability distribution of a Majorana particle most significantly. More specifically, the mass causes the shift in the expectation value of $x$, and a Majorana particle lie on the point with the highest shift value.

In addition, we show that the error of recovering the Dirac evolution is related to the mass term, and a Majorana fermion depends more sensitively on the accuracy of the approximation than a Dirac particle. To achieve the second order accuracy, we modify the general QW operator $\boldsymbol{W}_{SB}$ by decomposing the coin operator symmetrically. We notice that the modified QW operator $\boldsymbol{W}_{BSB}$ can be easily implement in the quantum circuit by change the rotation angle of the first coin operator from $\theta$ to $\frac{\theta}{2}$, and hence it is compatible with existing efficient QW algorithm. As a result, with this higher-order correction algorithm, we can derive a more precise result with limited QV of NISQ computer.

Certainly, Majorana fermion in 1+1-dimensional spacetime is not completely analogous to the case in 3+1-dimensional spacetime[55]. However, in both cases, the mass is defined as the coupling strength between the two chiralities (the $\psi_L$ and the $\psi_R$). The Dirac fermion mass term couples the two chiralities, while the Majorana one couples chiralities related by Majorana condition which couple the positive and negative energy states. In both (1+1) and (1+3)-dimensions, there are Majorana oscillations at zero momentum, but for non-zero momentum, (1+3)- dimension having four components instead of two components in (1+1)-dimension can be better handled. However, quantum computer realization may be more challenging. We will present more detailed analysis elsewhere. The DTQW algorithm for the near-term quantum computer have ability to unveil the relation between the chirality and mass, although the general DTQW initial state cannot represent real particle as wave packet. Because, in the DTQW algorithm, the motion of a particle is controlled by its chirality. Also, our results indicated that the modified DTQW operator reduce the fluctuation of the entanglement between coin space and position space and increase the entanglement when the state satisfied Majorana condition. These results provide the opportunity to tackle the fundamental physical problem by the quantum information processing with the NISQ computer.

Applying DTQW algorithm on an initial state of plane wave solution with $p = 0$, the phenomenon of Majorana oscillation is indeed observed as shown in **Fig. 3 (b)**. To use same DTQW methodology further on different initial states of $\Psi_M(0)$ and $\Psi_D(0)$ with $\langle p \rangle = 0$ separately, the time evolution are shown as **Fig. 4** and **Fig. 5** with walk operators with different order correction. Both the numerical and NISQ computer results showed the initial state, $\Psi_M(0)$, is more sensitive to the noises and errors than the initial state, $\Psi_D(0)$. Also, the mass term influence on the Majorana particle more than on the Dirac particle for the Majorana condition in our NISQ computer results. Further investigations of the response in (3+1)-dimensional spacetime for initial states of $\Psi_M(0)$ and $\Psi_D(0)$ deserved to study in details when a larger scale quantum computer available.

## METHODS

### Circuit implements detail

To simulate the DQW in a quantum computer, we use the IBM quantum computer: ibm_lagos which is 7 qubits processor and develop the algorithm using the open source SDK-Qiskit[56]. For sampling, the number of repetitions of each circuit is 8192. Specially, we execute the circuit of time evolution by Qiskit Runtime Service to increase the efficiency.

For the QW algorithm, the most complex part of the circuit is the conditional shift operator. In general, the conditional shift operator can be directly constructed by control-incrementors and control-decrementors[53]. However, the multiple-control gate of incrementors and decrementors increase circuit-depth enormously, making it hard to be realized in noisy intermediate-scale quantum (NISQ) computer.

Recently, researchers have shown that IBM quantum computers are able to produce non-trivial results for DTQW[57], and the strategy of reducing circuit depth have been developed[51,53,54]. In addition, since the mapping of the position state is not unique, we can reduce the circuit-depth by using specific mapping and fixing the initial state of the walk. In this work, we follow the strategy of Ref.[50,51] to fix the initial position in $x = 0$ and the position mapping is given in the **Table 1.** For seven-step walks, the corresponding circuit is shown in **Fig. 7** This symmetrical mapping reduces the position size from 16 to 15 ($|1000\rangle$ is truncated).

**Table 1 Mapping between the position states mapping and multi-qubits states.**

| $x$ | -7 | -6 | -5 | -4 | -3 | -2 | -1 | 0 | 1 | 2 | 3 | 4 | 5 | 6 | 7 |
|---|---|---|---|---|---|---|---|---|---|---|---|---|---|---|---|
| Qubits state | 1 | 1 | 1 | 0 | 0 | 0 | 0 | 0 | 0 | 0 | 0 | 1 | 1 | 1 | 1 |
| | 0 | 0 | 0 | 1 | 1 | 1 | 1 | 0 | 0 | 0 | 0 | 1 | 1 | 1 | 1 |
| | 0 | 1 | 1 | 0 | 0 | 1 | 1 | 0 | 0 | 1 | 1 | 0 | 0 | 1 | 1 |
| | 1 | 0 | 1 | 0 | 1 | 0 | 1 | 0 | 1 | 0 | 1 | 0 | 1 | 0 | 1 |
| | 0 | 1 | 1 | 0 | 0 | 1 | 1 | 0 | 0 | 1 | 1 | 0 | 0 | 1 | 1 |
| | 1 | 0 | 1 | 0 | 1 | 0 | 1 | 0 | 1 | 0 | 1 | 0 | 1 | 0 | 1 |

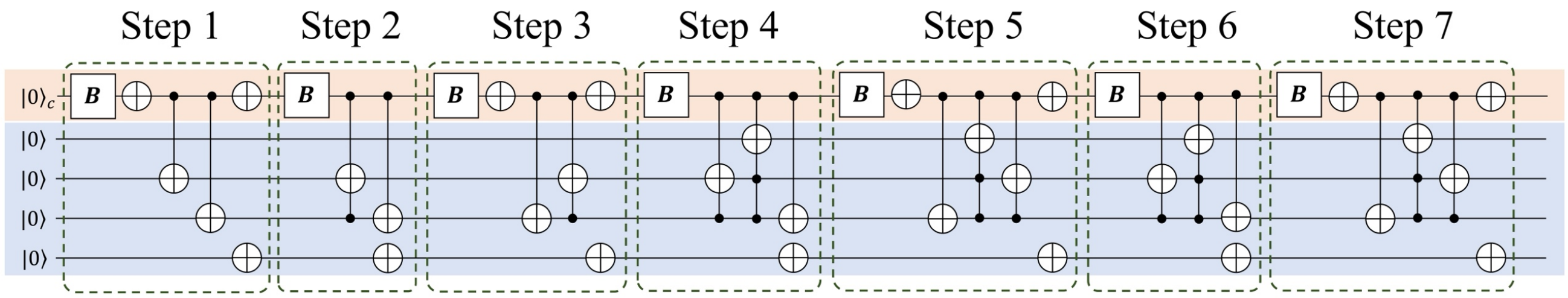

**Fig. 7 The Quantum circuit for real quantum device for the first seven steps on five qubit system.** The initial state is fixed to be $|0\rangle \otimes |x=0\rangle \equiv |0\rangle \otimes |0000\rangle$ for the position state mapping with the multi-qubit state as shown in **Table 1**.

## ACKNOWLEDGEMENTS

W.-T. W thanks the support of the NTU-IBM Q Hub at National Taiwan University, sponsored by the Ministry of Science and Technology, Taiwan, R.O.C. under Grant no. MOST 107–2627-E-002–001-MY3, 109-2627-M-002 -002, 110-2627-M-002 -003, and 111-2119-M-002 -006 -MY3. C.-R. C. thanks the support of the Quantum Information Center at Chung Yuan Christian University from the Ministry of Science and Technology, Taiwan, under grant No. MOST 109-2112-M-002-017-MY3 and 111-2119-M-033-001.

## AUTHOR CONTRIBUTIONS

W.-T. W. and C.-R. C. contributed to the major design and implementation of the research, to the analysis of the results, and to the writing of the paper. X.G. H. and H.-C. K. studied the theoretical analysis on quantum walk dynamics. X.G. H. developed the analytic formulation of higher-order error correction for DQW. All authors carefully reviewed the manuscript.

## REFEREBCES